\documentclass{article}

\usepackage{arxiv}

\usepackage[utf8]{inputenc} 
\usepackage[T1]{fontenc}    
\usepackage{hyperref}       
\usepackage{url}            
\usepackage{booktabs}       
\usepackage{amsfonts}       
\usepackage{nicefrac}       
\usepackage{microtype}      
\usepackage{lipsum}

\usepackage{graphicx}
\graphicspath{ {./images/} }
\usepackage{graphicx}
\usepackage{float}
\usepackage{amsmath}
\usepackage{dutchcal}
\usepackage{hhline}
\usepackage{tabularx,booktabs,color}
\usepackage{stfloats}
\usepackage{upgreek}
\usepackage{bm}
\usepackage{array}
\usepackage{booktabs}

\title{Photonic Convolution Neural Network Based on Interleaved Time-Wavelength Modulation}

\author{
 Yue Jiang, Wenjia Zhang*, Fan Yang, and Zuyuan He\\
 State Key Laboratory of Advanced Optical Communication Systems and Networks, \\
 Shanghai Jiao Tong University, Shanghai, China 200240\\
 \texttt{*wenjia.zhang@sjtu.edu.cn}
}

\begin{document}
\maketitle
\begin{abstract}
\vspace{-0.33cm}
Convolution  neural network (CNN), as  one  of the  most  powerful  and  popular  technologies, has achieved remarkable progress for image and video classification since its invention in 1989. 
However, with the high definition video-data  explosion, convolution layers in  the  CNN  architecture  will  occupy a great amount of  computing  time  and memory resources due to high computation complexity of matrix multiply accumulate  operation.
In this paper, a  novel  integrated  photonic  CNN  is  proposed  based on  double  correlation  operations  through  interleaved  time-wavelength modulation. 
Micro-ring based multi-wavelength manipulation and single dispersion medium are utilized  to realize convolution operation and replace the conventional optical delay lines.
200 images are tested in MNIST datasets with accuracy of 85.5$\%$ in our photonic CNN versus 86.5$\%$ in 64-bit computer.
We also analyze the computing error of photonic CNN caused by various micro-ring parameters, operation baud rates and the characteristics of micro-ring weighting bank.
Furthermore, a tensor processing unit based on \(4\times 4\) mesh  with 1.2 TOPS (operation per second when 100$\%$ utilization) computing capability at 20G baud rate is proposed and analyzed to form a paralleled photonic CNN. 
\end{abstract}


\section{Introduction}
As the driving force of Industry 4.0, artificial intelligence (AI) technology is leading dramatic changes in many spheres such as vision, voice and natural language classification \cite{lecun2015deep}. 
Convolution neural networks (CNN), as one of the most powerful and popular technologies, has achieved remarkable progress for image classification through extracting feature maps from thousands of images \cite{krizhevsky2012imagenet}.
In particular, CNN, with various structures such as AlexNet \cite{krizhevsky2012imagenet}, VGG16 (or 19) \cite{simonyan2014very} and GoogleNet \cite{Szegedy2014Going}, is mainly consisted of two parts: convolution feature extractors to extract the feature map through multiple cascaded convolution layers, and fully connected layers as a classifier.
In the CNN architecture, convolution layers will occupy most of computing time and resources \cite{Jia2014Learning} due to high computation complexity of multiply accumulate operation and matrices multiply accumulate operation (MMAC) \cite{He2015Convolutional}. Therefore, image to column algorithm combined with general matrix multiplication (GeMM) \cite{Dukhan2019The, liu2020systolic} and Winograd algorithms \cite{Lavin2015Fast} were proposed to accelerate the original 2-D convolution operation (2Dconv) 
due to the improvement of memory efficiency \cite{Cho2017MEC}. With the high definition video-data explosion, algorithm innovation can not achieve outstanding performance gain without hardware evolution. Therefore, innovative hardware accelerators have been proposed and commercialized in the forms of application specific integrated circuit (ASIC) \cite{Luo2016DaDianNao}, graphics processing unit (GPU) \cite{suita2019efficient,Suita2020Efficient} and tensor processing unit (TPU) \cite{jouppi2017datacenter}. 
However, it has become overwhelmed for conventional electronic computing hardware to adapt the continuedly developing CNN algorithm \cite{wang2018neural}.

In the meantime, integrated photonic computing technology presents its unique potential for the next generation high performance computing hardware due to its intrinsic parallelism, ultrahigh bandwidth and low power consumption \cite{caulfield2010future}. Recently, significant progress have been achieved in designing and realizing integrated optical neural networks (ONN) \cite{shen2017deep,ma2020photonic,tait2018feedback}.  
The fundamental components including Mach-Zehnder interferometers (MZI) \cite{ma2020photonic}  and micro-ring
resonators (MRR) \cite{tait2018feedback} have been widely employed to compose a optical matrix multiplier unit (\(\rm{OM^2U}\)), which is used to complete the MMAC operation. 
In order to construct full CNN architecture, electrical control unit like field programmable gate array (FPGA) is required to send slices of input images as voltage control signals to optical modulators and also operate nonlinear activation.
For instance, an \(\rm{OM^2U}\) controlled by FPGA, has been proposed by using fan-in-out structure based on microring resonators \cite{cheng2020silicon}. 
Similarly, CNN accelerator based on Winograd algorithm in the work of \cite{mehrabian2019winograd} is also composed of an \(\rm{OM^2U}\) based on MRR and electronic buffer.
However, the proposed photonic CNN architecture controlled by electronic buffer rely on electrical components for repeatedly accessing memory to extract the corresponding image slices (or slice vectors) and are finally constrained by memory access speed and capacity. 
In 2018, photonic CNN using optical delay line to replace the electronic buffer was firstly proposed in \cite{bagherian2018chip}. Based on the similar idea, the researchers have developed an optical patching scheme to complete the 2-D convolution in \cite{xu2020optical}, where the wavelength division multiplexing (WDM) method is used\cite{bagherian2018chip}.

In our previous work \cite{huang2019programmable}, wavelength domain weighting based on interleaved time-wavelength modulation was demonstrated to complete the MMAC operation. The idea of multi-wavelength modulation and dispersed time delay can realize matrix vector multiplication by employing time and wavelength domain multiplexing. However, the cross-correlation operation between an input vector and a single column of weighting matrix is operated through sampling process by generating a large amount of useless data. Moreover, a 2Dconv operation can be decomposed as the sum of multiple double correlation operations between vectors.
In this paper, a novel integrated photonic CNN is proposed based on double correlation operation through interleaved time-wavelength modulation. 
Microring based multi-wavelength manipulation and single dispersion medium are utilized  to realize convolution operation and replace the conventional optical delay lines used in \cite{bagherian2018chip} and \cite{xu2020optical}.
200 images are tested in MNIST datasets with accuracy of 85.5$\%$ in our PCNN versus 86.5$\%$ in 64-bit computer. We also analyze the error of PCNN caused by high baud rate and the characteristics of MRR weighting bank. Furthermore, a tensor processing unit based on \(4\times 4\) \(\rm{OM^2U}\) mesh  with 1.2 TOPS  (operation per second when 100$\%$ utilization) computing capability at 20G baud rate for MZM architecture is proposed and analyzed to form a paralleled photonic CNN.

\section{Physical Implementation of the OCU}
\subsection{Optical Convolution Unit}
The convolution layer is the key building block of a convolution network that operates most of the computational heavy lifting. Convolution operation essentially performs dot products between the feature map and local regions of the input. This operation will be iterated in the input image at stride of given location along both width and height. Therefore, the designed operation will consume a lot of memory, since some values in the input volume are replicated multiple times due striding nature of this process.  

\begin{figure}[t] 
\begin{center} 
\includegraphics [width=1\textwidth]{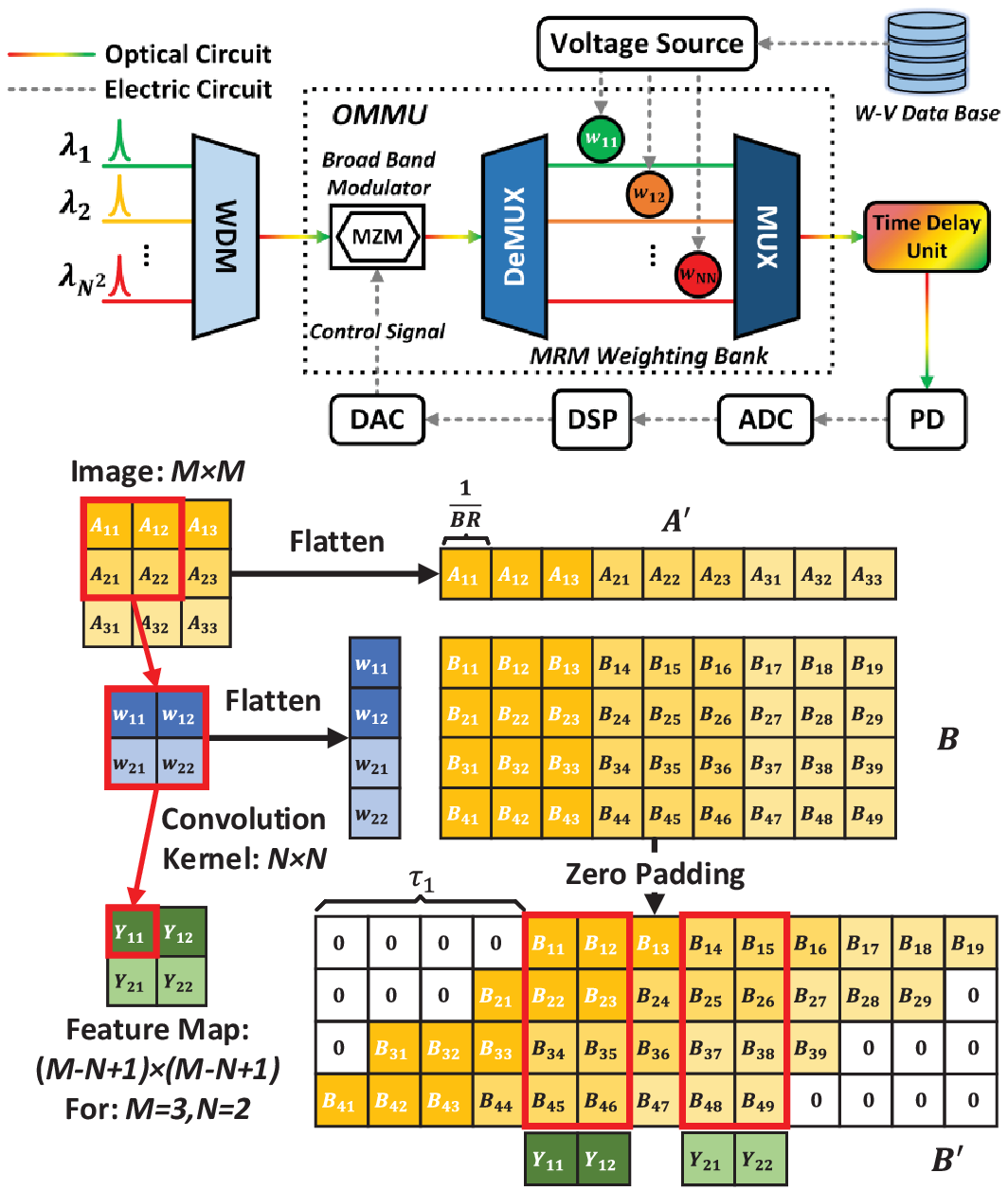} 
\caption{(a) Structure of the OCU, where the 2Dconv operation shown in (b) is done. MZM: Mach Zehnder modulator, W-V Data Base: set up following the process shown in Fig. \ref{fig:2}(b) to generate the voltage control signal loaded on the MRR weigthting bank, PD: Photodetector to covert optical signal into electric domain, ADC and DAC: 
Analog-to-Digital and Digital-to-Converter respectively, DSP: Digital signal processing where the sampling, nonlinear, and pooling operation is done.}  
\label{fig:1}
\end{center}

\end{figure}

In the proposed photonic CNN as shown in Fig. \ref{fig:1}(a), the optical convolution unit (OCU) is consisted of  \(\rm{OM^2U}\) and dispersed time delay unit (TDU). The single 2Dconv operation for the  \(M \times M\) input image A and \(N \times N\) convolution kernel w is executed during one period in the OCU, which can be written as:
\begin{equation}
    {Y_{m,n}} = \sum\limits_{i = 1}^N {\sum\limits_{j = 1}^N {({w_{i,j}} \cdot {A_{m + i - 1,n + j - 1}})} } 
   \label{Eq:1}
\end{equation}

\begin{figure*}[t] 
\includegraphics [width=1\textwidth]{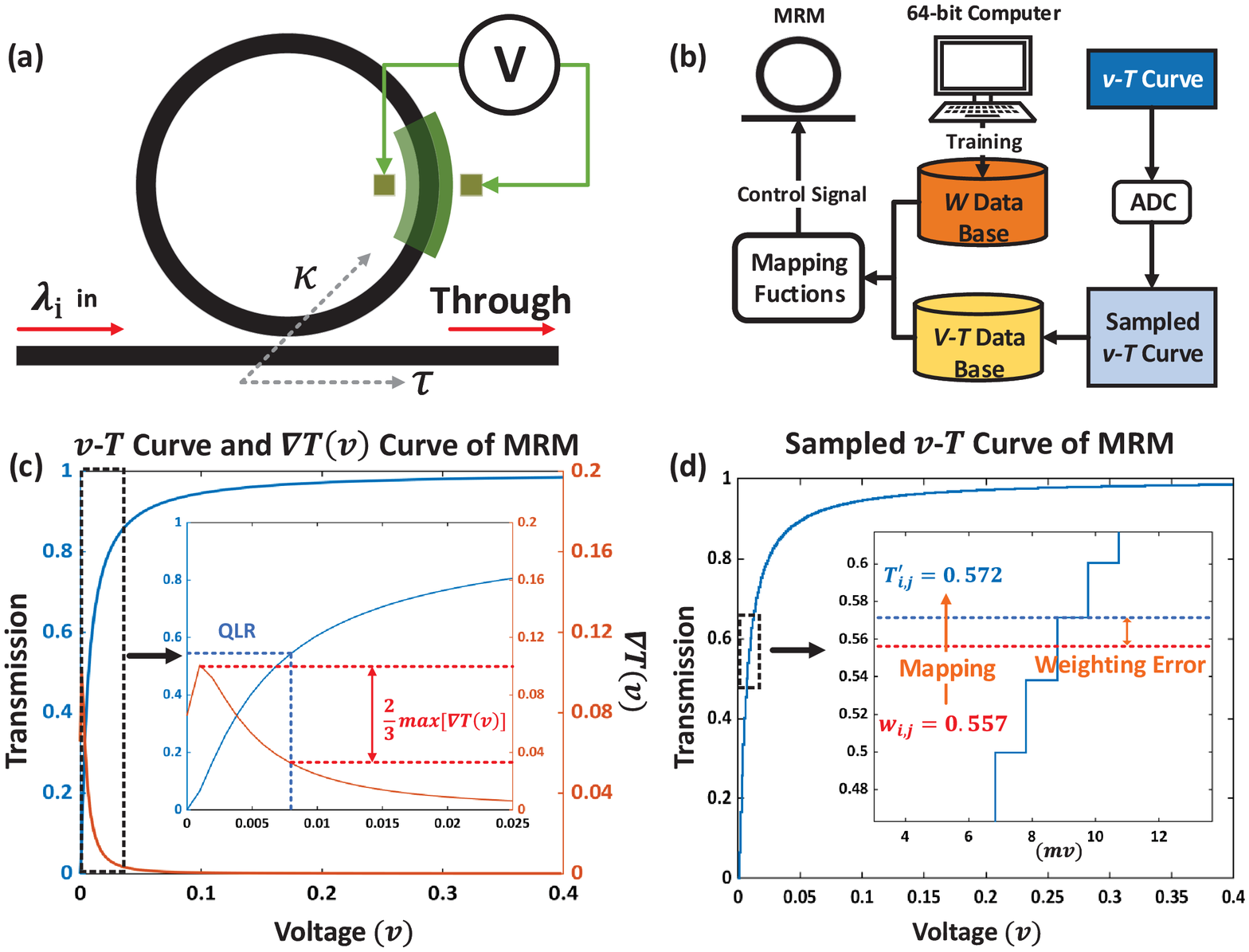} 
\caption{ (a) Schematic of MRR based on EO effect, (b) Mapping process of \(\rm{w}\) to \(T-V\). (c) \(v-T\) and \(\nabla T(v)\) curve of MRR, the QLR (quasi-linear  region)   in this paper is defined as the region between 0 \(v\) and the corresponding voltage at the highest \(1/3\) of the \(\nabla T(v)\) curve , (d) \(v-T\) curve sampled by ADC with 10-bit precision, note that there are error \(w'\) existed between theoretical mapping points \(w_{i,j}\) and true mapping points \(T'_{i,j}\).}  
\label{fig:2}
\end{figure*}

Here we set \(M=3\), \(N=2\) for example in Fig. \ref{fig:1}(b), the input image A is flattened into a normalized \(1\times{M^2}\) vector \(\rm{A}'\) which is modulated by a MZI modulator on multi-wavelength optical signals with \(N^2\) wavelengths: \({\lambda _1}\), \({\lambda _2}\)... \({\lambda _{{N^2}}}\) at certain Baud Rate (marked as \(BR\) in equations). The intensity of each frequency after modulation, \({I_{A'}}(t)\) can be written as

\begin{equation}
\begin{array}{l}
{I_{A'}}(t) = \sum\limits_{l = 1}^M {\sum\limits_{k = 1}^M {{I_{input}} \cdot {A_{l,k}}} }  \cdot Square(t)\\
Square(t) = U[t - \frac{{(l - 1) \times M + k}}{{BR}}] - U[t - \frac{{(l - 1) \times M + k + 1}}{{BR}}]
\label{Eq:2}
\end{array}
\end{equation}
Where the \(U(t)\) is the step function, and the \(I_{input}\) is the intensity of a single channel in WDM sources, which are equal for all frequencies. 
Optical signals of different wavelengths are separated by the DEMUX, and sent to the corresponding MRRs. There are \(N^2\) MRRs \(R_1\),\(R_2\), …, \(R_{N^2}\) compose as a MRR weighting Bank. The transmission (\({T_{(i - 1) \times N + j}}\)) of each MRR are set to the \(w_{i,j}\) and tuned by the voltage bias from voltage source or an arbitrary waveform generator. 
The control signal is generated from the w-V database which stores the mapping between the \(\rm{w}\) and \(\rm{V}\). The output intensity of each MRR \(\mathit{I_{R_{(i-1)\times N+j}}(t)}\) with circuit time delay \({\tau _c}\) can be written as

\begin{equation}
\mathit{I_{R_{(i-1)\times N+j}}(t)}=\mathit{I_{{A}'}(t-\tau _{c})}\cdot w_{i,j}
\label{Eq:3}
\end{equation}
Optical signals of different wavelengths are combined as the matrix \(\rm{B}\) shown in Fig. \ref{fig:1}(b) in time domain, by passing through the MUX. The output intensity \({I_{OM^2U}(t)}\) of the \(\rm{OM^2U}\) with the time delay \({\tau_c}'\) is
\begin{equation}
\mathit{I_{OM^2U}}(t)=\sum_{i=1}^{N}\sum_{j=1}^{N}\mathit{I}_{{A}'}(t-\tau _{c}')\cdot w_{i,j}
\label{Eq:4}
\end{equation}
Which is equal to the MMAC operation between  the flattened convolution kernel vector \(\rm{w}'\) and the matrix \([\rm{A}'^T,... , \rm{A}'^T]\)which contains \({N^2}\) copies of \(\rm{A}'\). As depicted in Fig. \ref{fig:1}(b), to complete the 2Dconv operation between \(\rm{A}\) and \(\rm{w}\), the corresponding elements in (\ref{Eq:1}) should be in the same column of the matrix \(\rm{B}'\), 
which can be realized by introducing different time delay \({\tau}_{(i-1)\times N+j}\) for wavelength \(\lambda_{(i-1)\times N+j}\)  in TDU to complete the zero padding operation:

\begin{equation}
\tau_{(i-1)\times N+j}=[(N-i)\times M)+N-j]/BR
\label{Eq:5}
\end{equation}
The intensity of the light wave passing through the TDU with the wavelength independent circuit time delay \(\tau_c''\) can be written as
\begin{equation}
\mathit{I_{TDU}(t)}=\sum_{i=1}^{N}\sum_{j=1}^{N}\mathit{I}_{A'}(t-\tau_c''-\tau_{(i-1)\times N+j})
\label{Eq:6}
\end{equation}
When optical signal is received by the photo-detector (PD), the \(I_{TDU}(t)\) convert to  \(V_{PD}(t)\). Refer to (\ref{Eq:6}), there are \(M^2+(N-1)\times (M+1)\) elements in each row of matrix \(\rm{B}'\), and the \(q^{th}\) column of which occupies one time slice in \(V_{PD}(t)\): from \(\tau_c''+(q-1)/BR\) to \(\tau_c''+q/BR\), compare the (\ref{Eq:1}) and (\ref{Eq:6}), when
\begin{equation}
\textit{q}=(M-N+1)\times (m-1)+(M+m)+n
\label{Eq:7}
\end{equation}
Where \(1\leq m,n\leq M-N+1 \), and set a parameter \(\sigma\) between 0 and 1, we have:
\begin{equation}
\mathit{Y}_{m,n}=V_{PD}[(t-\tau_c''-q+\sigma)/BR]
\label{Eq:8}
\end{equation}
When \(M=3\), \(N=2\) shown in Fig. \ref{fig:1}(b), the sum of \(\rm{B}_{i,5}'\), \(\rm{B}_{i,6'}\), \(\rm{B}_{i,8}'\), and \(\rm{B}_{i,9}'\) corresponding to \(Y_{1,1}\), \(Y_{1,2}\), \(Y_{2,1}\), and \(Y_{2,2}\), respectively. A programmed sampling function refer to (\ref{Eq:7}) and (\ref{Eq:8}) is necessary in digital signal processing, and the parameter \(\sigma\) decides the position of optimal sampling point, which needs to be adjusted at different bit rates. According to the (\ref{Eq:5}), the row \(\rm{B}'_q\) of matrix \(\rm{B}'\) can be divided into \(N\) groups with \(N\) vectors composed as a matrix of \(\rm{Group}_{i,j}=\rm{B}'_{(i-1)\times N+j}\) , where \(i,j \leq N\). The kernel elements multiplied with vector \(\rm{A}'\) in \(\rm{Group}_i\) are \([w_{i,1}, w_{i,2}, ..., w_{i,N}]\), which are the elements in the same row of a convolution kernel \(w\). Refer to (\ref{Eq:5}), the difference of the time delay in between two adjacent rows in the same group is equal to \(1/BR\), whereas the difference of time delay between \(\rm{Group}_{i,j}\) and \(\rm{Group}_{i+1,j}\) is equal to \(M/BR\). The sum of \(q^{th}\) column in the same group of \(\rm{B}'\) can be written as
\begin{equation}
\sum \rm{Group}_{i}(q)=\sum_{j=1}^{N}w_{i,j}\cdot A'_{q+j-N}
\label{Eq:9}
\end{equation}
which is actually the expression of the cross-correlation (marked as \(R(x,y)\)) between vector \([w_{i,1}, w_{i,2}, ..., w_{i,N}]\) and \(\rm{A}'\). Therefore, the 2Dconv operation can be decomposed as the sum of multiple double correlation operation between vectors as follows

\begin{equation}
\sum_{p=1}^{N^2}\rm{B}'_{p}=\sum_{i=1}^{N}R[R(\rm{A}',\rm{w}_i),Flatten(\rm{C}_i)]
\label{Eq:10}
\end{equation}
where \(\sum_{i=1}^{N}\rm{C}_i\) is an identity matrix with the size of \(N\times N\), and the elements at the \(i^{th}\) row and column of \(\rm{C}_i\) is equal to 1, the other elements equal to  0. The matrix  \(\rm{C}_i\) is flattened in to a \(1 \times N^2\) vector, and  cross-correlation operation is denoted as \(R(\rm{A}',\rm{w}_i)\).

\subsection{The mapping of weight elements to voltage}

The MRRs based on electro-optic or thermal-optic effect are used in weighting Bank of OCU. Refer to (\ref{Eq:3}), the elements of convolution kernel \(w_{i,j}\), trained by 64-bit computer, are usually normalized from 0 to 1, which needs to be mapped into the transmission of MRRs. As shown in Fig. \ref{fig:2}(a), according to \cite{tazawa2006ring,bortnik2007electrooptic}, the transmission of the through port of MRR based on electro-optic effect is tuned by voltage bias \(V\) loaded on the electrode of MRR, which can be written as: 
\begin{equation}
T=1-\frac{(1-\alpha ^2)(1-\tau ^2)}{(1-\alpha\tau)^2+4\alpha\tau\sin^2(\theta /2) },
\theta=\theta_0+\pi V/V_{\pi}
\label{Eq:11}
\end{equation}
Where \(\tau\) is the amplitude transmission constant between the ring and the waveguide, \(\alpha\) is the round-trip loss factor, and \(\theta\) is the round-trip phase shift, \(\theta_0\) is the bias phase of the MRR, and \(V_{\pi}\) is the voltage loaded on the MRR when \(\theta=\pi \) , which is decided by the physical parameters of the waveguide. 
The curve of V-T is shown in Fig. \ref{fig:2}(c). A voltage source with specific precision (10-bit in our evaluation) sweeps the output voltage with the minimum step from 0 to 0.4, which is loaded on the MRR. The transmission metrics of MRR at different voltages are recorded accordingly. 
As shown in Fig. \ref{fig:2}(d), the processing actually equivalent to sampling the curve of V-T by using an analog-to-digital converter (ADC) with same precision of the voltage source.  
If \(\left | w_{i,j} \right |\leq 1\), \(w_{i,j}\) can be mapped directly into \(T\), 
the weighting voltage \(V\) can be figured out by searching the number which is closest to \(w_{i,j}\) in the  database T-V.
Otherwise, the whole convolution kernel should be normalized through being divided by the max of \(w_{i,j}\). Then, the normalized \(\rm{w}_{nor}\) matrix is utilized to control signal matrix \(\rm{V}\).
Another mapping method is designed by using part of quasi-linear region in V-T curve of MRR, where the matrix \(\rm{w}\) needs to be normalized by multiplying \(max(\rm{T}_{linear})/max(\rm{w})\). 
Note that the error weighting error occurs during the mapping process as shown in Fig. \ref{fig:2}(d). There will be a difference \(w'\) between the actual transmission of MRR \(T'\) and an ideal mapping point \(T\). So the weighting error and outcome of the \(\rm{OM^2U}\),  \(\rm{Y}'\) can be written as (\ref{Eq:12}), where \(\rm{Y}\) is the theoretical outcome of the \(\rm{OM^2U}\), and \(\rm{Y'} \rightarrow \rm{Y}\) when \(\rm{w}' \rightarrow \rm{0}\).

\begin{equation}
\begin{array}{l}
\rm{w}' = \rm{T}' - \rm{T}\\
\rm{Weighting\,Error} = [{{\rm{A}}'^T},...,{\rm{A}}'^T] \times \rm{w}'\\
\rm{Y} = [{\rm{A}'^T},...,{\rm{A}'^T}] \times (\rm{w}+\rm{w}')\\
\rm{Y}'= \rm{Y} + \rm{Weighting \,Error}
\end{array}\
\label{Eq:12}
\end{equation}

\subsection{Dispersed Time Delay Unit}
\begin{figure}[t] 
\begin{center} 
\includegraphics [width= 0.75\textwidth]{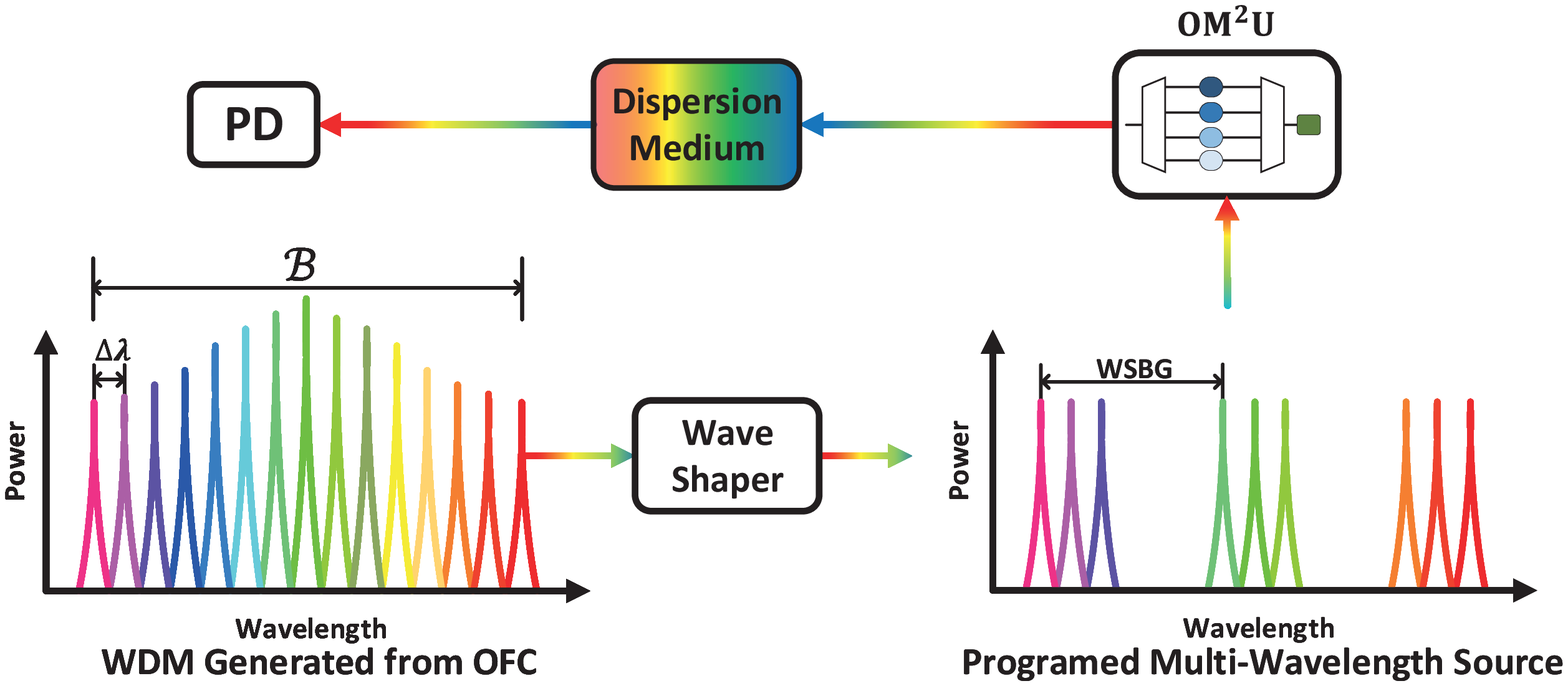} 
\caption{TDU based on single dispersion medium and Programmed multi-wavelength source, which is generated by the optical comb and wave shaper, with \(N\) groups wavelengths, and \(N\) wavelengths in each group, with the wavelength distance of \(\Delta \lambda\), and the wavelength space between adjacent groups marked as \(WSBG=\Delta \lambda \cdot M\).} 
\label{fig:3}
\end{center} 
\end{figure}

The zero padding operation is executed  by offering different time delay for each channel of multi-wavelength light source in time delay unit.
In our previous work \cite{huang2019programmable}, the \(\rm{OM^2U}\) based on wavelength division weighting method with single dispersion compensating fiber (DCF) was proposed, where the correlation operation between two vectors is realized in time domain refer to (\ref{Eq:9}.)
Based on the \(\rm{OM^2U}\) in \cite{huang2019programmable}, the TDU can be implemented with single dispersion media combined with programmed multi-wavelength light source (PMWS) shown in Fig. \ref{fig:3}, which can be generated by a shaped optical frequency comb refer to (\ref{Eq:5}). 
The programmed light source contains \(N\) groups wavelengths, and \(N\) wavelengths are included in each group with the wavelength spacing of \(\Delta \lambda\), the wavelength spacing between adjacent groups is equal to \(M\times \Delta \lambda\). The requirements of programmed multi-wavelength light source can be written as

\begin{equation}
\left\{\begin{matrix}
 PMWS_{i,j}- PMWS_{i,j-1}=\Delta \lambda\\
 PMWS_{i,j}- PMWS_{i-1,j}=M\times\Delta\lambda 
\end{matrix}\right.
\label{Eq:13}
\end{equation}
where \(PMWS\) is programmable multiple-wavelength source, which is sent to the dispersion media with length of \(L\) (km), and the dispersion of \(D\) (s/nm/km). Therefore,  the time delay difference marked as TDD in \ref{Eq:14}) are introduced for optical signal with wavelength \( PMWS_{i,j}\) to the \( PMWS_{1,1}\). This value is equal to

\begin{equation}
    TDD_{i,j}=( PMWS_{i,j}- PMWS_{1,1})\times LD\\
\label{Eq:14}
\end{equation}

When \(TDD_{i,j}-TDD_{i,j-1}=1/BR\), (\ref{Eq:14}) is equivalent to (\ref{Eq:5}), i.e. zero padding operation is conducted when multi-wavelength signals passing through the dispersion media.  Note that there exist challenging tasks in implementing the TDU structure as shown in Fig. \ref{fig:3}. It is essential to design the frequency comb with large enough number and density of lines combine with dispersion media with flat, large enough  \(D\) (s/nm/km) and low loss.
The bandwidth, \(\mathcal{B}\) with the number of lines, \(\mathcal{k}\), and the length of DCF, \(L\) needed can be calculated as:
\begin{equation}
\left\{\begin{matrix}
\mathcal{B}=(M+1)\times (N-1)\times \Delta  \lambda\\ 
\mathcal{k}=\mathcal{B}/\Delta\lambda +1\\ 
L=(BR\times D\times \Delta\lambda)^{-1}
\end{matrix}\right.
\label{Eq:16}
\end{equation}
 
In this paper we take frequency comb with \(\Delta\lambda\approx 0.2 \) nm as reported in \cite{Liu2020Author} and DCF (suppose \(D\) is flat for all wavelength) with \(D=-150\) (ps/nm/km), to perform MNIST handwritten digit recognition
task, where \(M=28\), \(N=3\) for example, refer to (\ref{Eq:16}) with \(\mathcal{B}=11.6\) nm, \(\mathcal{k}=59\) lines, and \(L\) = 1.67 km at \(BR =20\) G. \\

Another widely discussed structure of dispersed delay architecture is based on multi-wavelength source and arrayed fiber grating, 
%
where the PMWS is not necessary, and the cost of source and bandwidth is much cheaper. However, at least \(N^2\) SMF are needed, which makes it hard to control the time delay of each wavelength precisely. \(N^2\) tunable time delay units for short time delay such as Fiber Bragg Grating  and \(Si_3N_4\) waveguide can be employed with proper delay controller to compensate the time delay error in each channel caused by fabrication process. Furthermore, the size of input images \(M_l\) for the \(l^{th}\) convolution layer is equal to half of \(M_{l-1}\) after pooling operation with stride of 2, the length of SMF for \(l^{th}\) convolution layer need to be adjusted according to \(M_l\), whereas the TDU based on PMWS and single DM can regulate the time delay with high robustness by reprogramming WDM source according to (\ref{Eq:14}).

\begin{figure*}[h]
\begin{center} 
\includegraphics [width=1\textwidth]{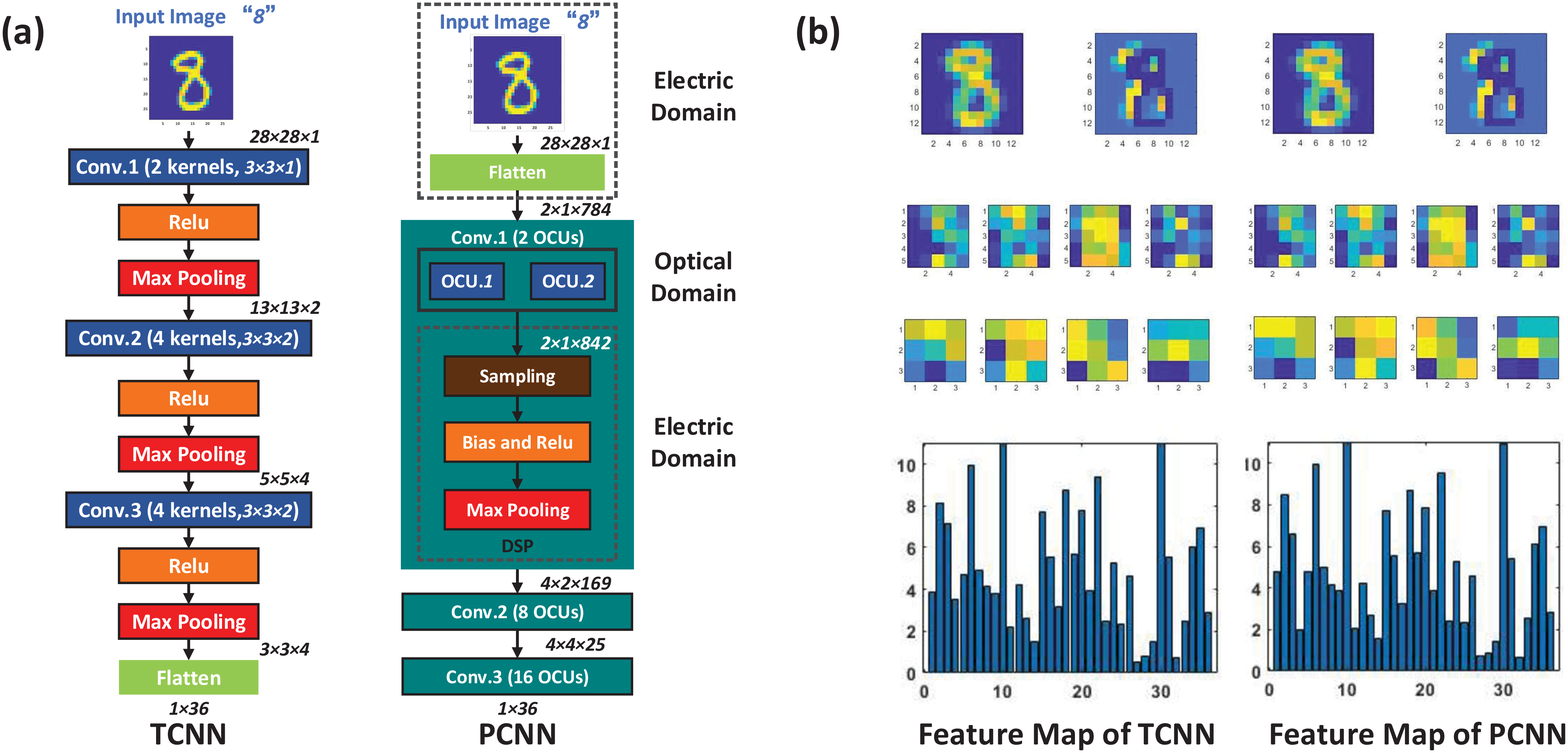} 
\caption{ (a)The architecture of convolutional neural network in TensorFlow (TCNN) with 3 convolution layers and the PCNN with the same architecture of TCNN, (b) Compare of feature map extracted by TCNN and reshaped feature vector extracted by PCNN. }  
\label{fig:4}
\end{center} 
\end{figure*}


\section{ Photonic CNN Architecture}
\begin{figure*}[h]
\begin{center} 
\includegraphics [width=1\textwidth]{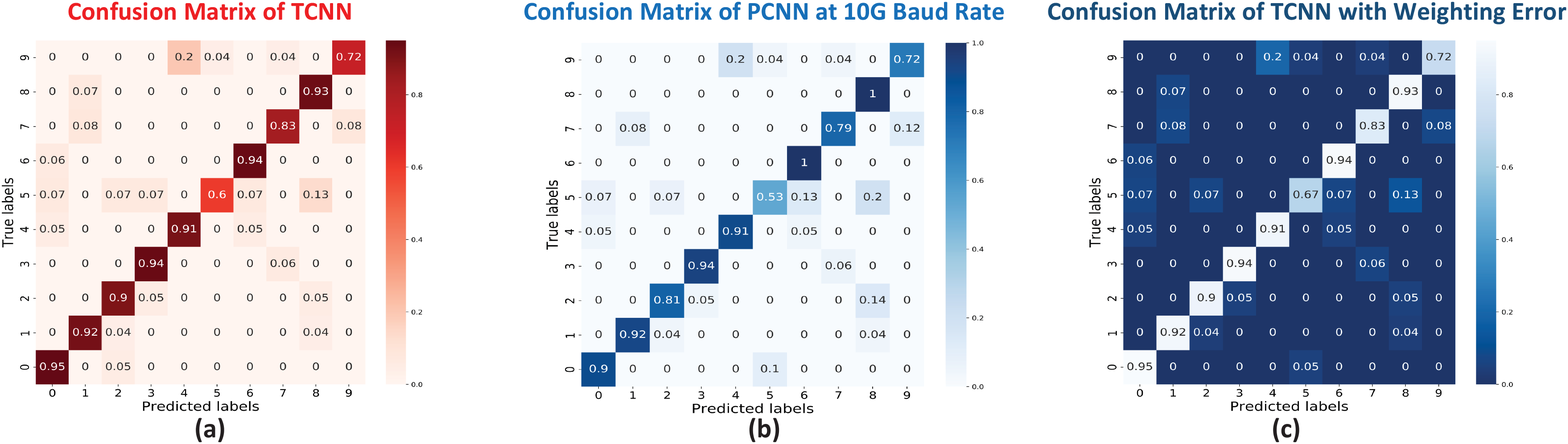} 
\caption{ (a) Confusion matrix of TCNN for 200 samples test, (b) Confusion matrix of PCNN at 10G Baud Rate, (c) Confusion matrix of TCNN with weighting bank error separated from PCNN.   }  
\label{fig:5}
\end{center} 
\end{figure*}

As shown in Fig. \ref{fig:4}(a), a simplified AlexNet convolution neural network for MNIST handwritten digit recognition task is trained offline on 64-bit computer in TensorFlow framework (TCNN), which is composed of 3 convolution layers, and 2 kernels \((3\times\ 3\times\ 1)\), 4 kernels \((3\times\ 3\times\ 2)\) and 4 kernels \((3\times\ 3\times\ 4)\) in the \(1^{st}\), \(2^{nd}\) and \(3^{th}\) convolution layer, respectively. The size of samples in MNIST written digital dataset \(28 \times 28 \times  1\) \((Width \times Height\times Channel) \), and the output shape for each layer is \((13\times\ 13\times 2)\), \((5\times 5\times 4)\), \((3\times 3\times 4)\), and finally a \((1\times 36)\) flatten feature vector (marked as FFV in equations) is output by the flatten layer. A PCNN simulator with the same architecture is set up based on Lumerical and Matlab to implement the optical domain and DSP part of the OCU. The \(V-T\) database is established by recording the transmission of corresponding wavelength at through port of the default MRR offered by lumerical, while sweeping voltage bias from 0 to 1.2 \(V\) with precision of 10-bit.
Then the mapping process shown in Fig. \ref{fig:2} is conducted to load convolution kernel into the PCNN simulator. 
The feature map extracted at each convolution layer of input figure “8” from TensorFlow and reshaped feature vector of PCNN are compared in Fig. \ref{fig:4}(b), which shows the feature map extraction ability of the PCNN. 
Finally 200 test samples in MNIST are extracted randomly and sent to the PCNN for test with the test accuracy is 85$\%$ at 10 G Baud Rate. Note that the TensorFlow is a simplified AlexNet whose classification accuracy for the same 200 test samples is only 86.5$\%$ in our 64-bit computer. 
The  confusion matrices of TensorFlow and PCNN at 10G Baud Rate are shown in Fig. \ref{fig:5} (a) and (b), respectively.

\section{Evaluation of Photonic CNN}
\subsection{Weighting Error of MRR Weighting Bank}
Equation (\ref{Eq:12}) shows that the weighting error occurs during mapping process, which is depending on the mapping precision \(P(v_i)\) of the MRR weigting bank. The \(P(v_i)\) can be evaluated by the difference of the \(T(v_i)\) \cite{cheng2020silicon}, which is

\begin{equation}
P({v_i}) = {\log _2}{[\nabla T({v_i})]^{ - 1}} = {\log _2}{[T({v_i}) - T({v_{i - 1}})]^{ - 1}}
\label{Eq:17}
\end{equation}

\begin{figure}[t]
\begin{center} 
\includegraphics [width=0.75\textwidth]{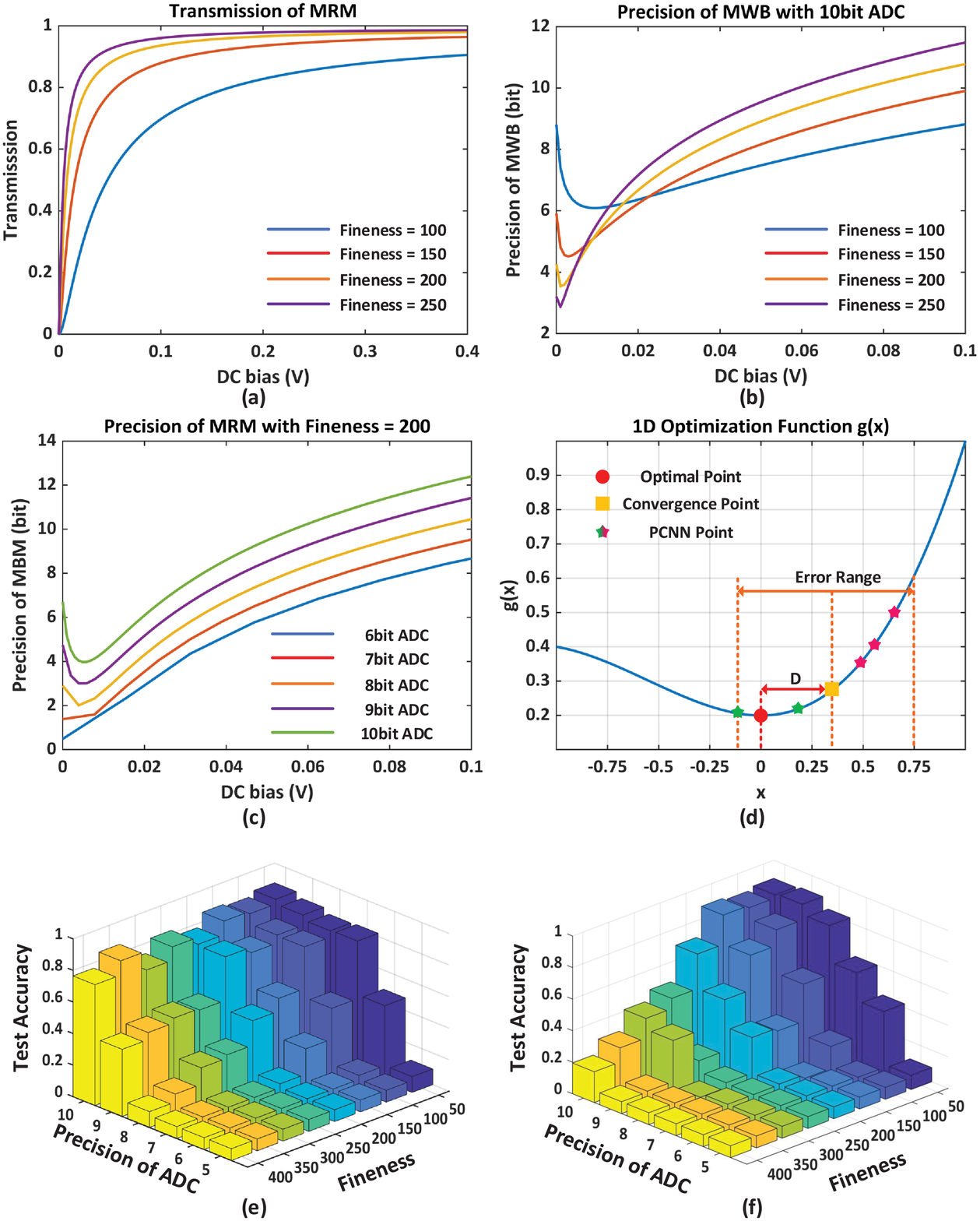} 
\caption{(a) \(T-V\) curve of MRR with different Fineness from 100 to 250, (b) Compare of weighting precision of MRR with different Fineness, (c) Compare of weighting precision of the MRR at different level of ADC precision, (d) The PCNN point which is equal to Convergence Point of TCNN with error may have shorter distance to the optimal point compared with that of TCNN, which leads to higher test accuracy, (e) Test Accuracy compare of MRR with different Fineness at distinct ADC precision level when \(w_{i,j}\) is mapped into \(T\) from 0 to 1, whereas (f) \(w_{i,j}\) is mapped into \(T\) in  quasi-linear region.}  
\label{fig:6}
\end{center} 
\end{figure}

As shown in Fig. \ref{fig:6}, we numerical analyze the \(P(v_i)\) of MRR with different fineness at distinct ADC precision level refer to (\ref{Eq:11}) and (\ref{Eq:17}).
In Fig. \ref{fig:6}(b), the MRR with smaller fineness has higher \(P(v_i)\) in  quasi-linear  region  (\(v_i\leq v_l\), where \(v_l\) is the boundary of  quasi-linear  region ). 
However, when \(v_i\geq v_l\), \(P(v_i)\) increases with the fineness. 
The precision of ADC also has impact on the \(P(v_i)\) of MRR. As depicted in Fig. \ref{fig:6} (c), \(P(v_i)\) increases with the precision of ADC. 
The weighting error separated from the PCNN is added to the flatten feature vector extracted from the TensorFlow CNN. 
The test accuracy of flatten feature vector is 87$\%$, with the confusion matrix  shown in Fig. \ref{fig:5} (c). 
Note that the test accuracy of flatten feature vector with error is higher than that in TensorFlow, the handwritten digital recognition task in this paper is a 36-dimensions optimal task. Here we use 1-dimension optimal function \(g(x)\) to explain. 
As shown in Fig. \ref{Eq:6}(d), there is a distance \(D\) between the optimal point and the convergence point of TensorFlow. The convergence  point of PCNN can be treated as optimal point of TCNN added with noises in error range. This deviation will probably lead to a closer location to the optimal point and therefore  a higher test accuracy with a certain probability.  The test accuracy of MRR with different fineness at distinct ADC precision level is shown in Fig. \ref{fig:6}(e), 
where the \(w_{i,j}\) is mapped into \(T\) from 0 to 1, whereas  \(w_{i,j}\) is mapped into \(T\) in  quasi-linear  region  in Fig. \ref{fig:6}(f). By comparing two figures, the MRR with low fineness and high ADC precision level  are preferred in high-speed photonic CNN.

\begin{figure}[t]
\begin{center} 
\includegraphics [width= \textwidth]{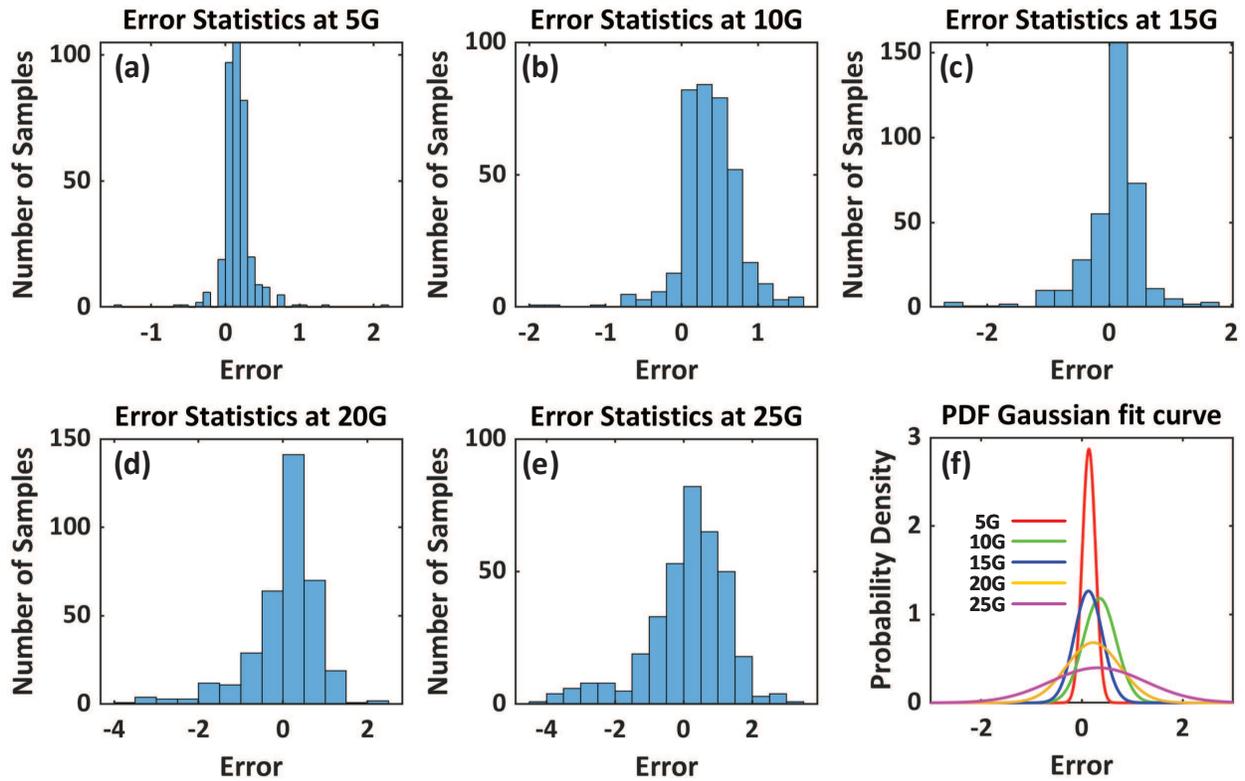} 
\caption{(a) to (e), the distribution statistics of \(\rm{Error}\) at the Baud Rate of 5,10,15,20, and 25G, respectively, (f) The Gaussian fit curve of probability density function (PDF) of \({\rm{Error}}\) at different Baud Rate. 
     }  
\label{fig:7}
\end{center} 
\end{figure}

 \begin{figure}[t]
\begin{center} 
\includegraphics [width=0.5\textwidth]{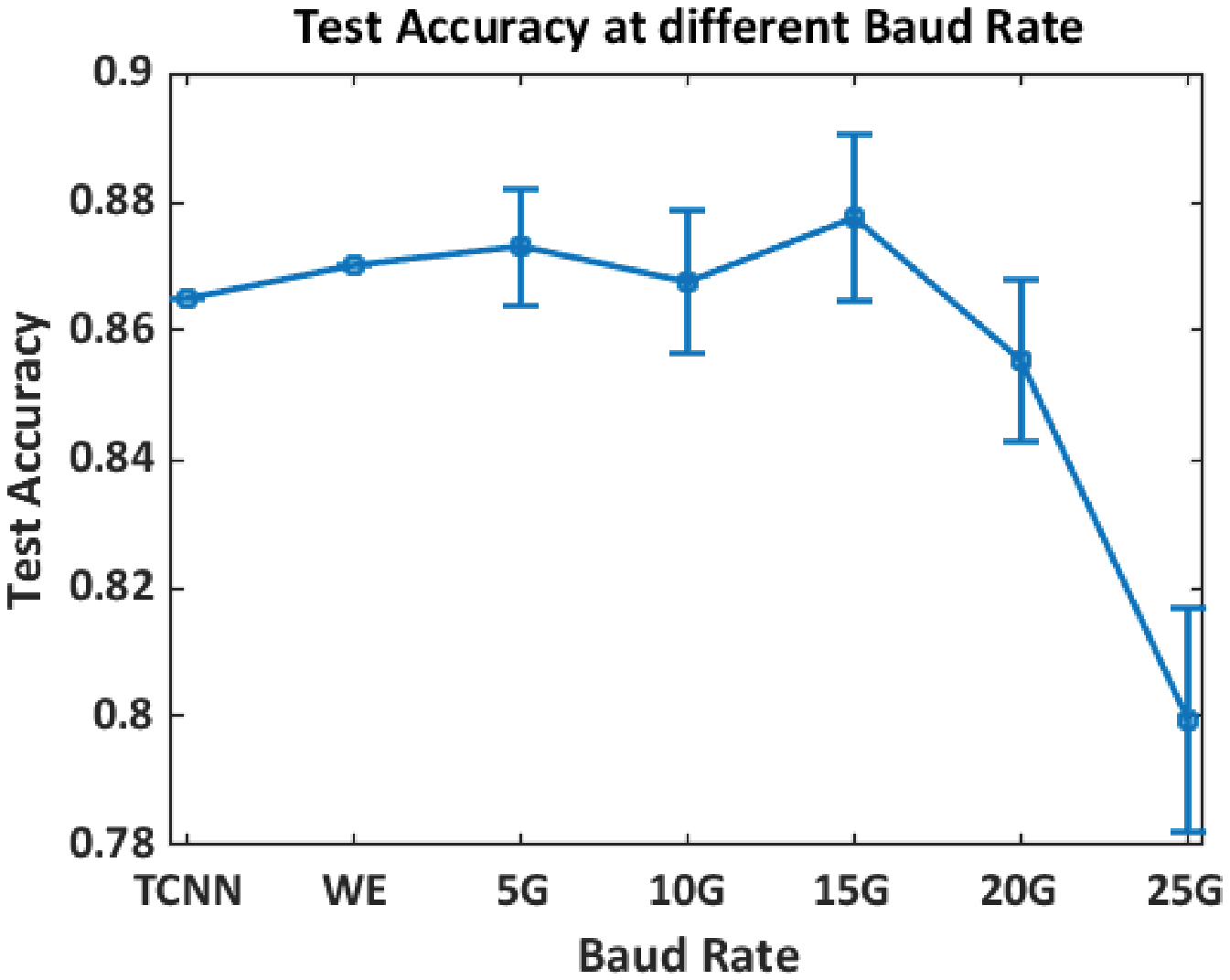} 
\caption{Performance of PCNN at different Baud Rate, the standard deviation is adopted here, note that,  \(\rm{Error}\) of TCNN and TCNN with Weighting Error (WE) are equal to \(\rm{0}\), i.e. the std at TCNN and Weighting Error are 0.   }  
\label{fig:8}
\end{center} 
\end{figure}
\subsection{Computation Speed}
The distortion will be introduced when high bandwidth signals passing through filters such as MRR. Moreover, the quantization noise for high frequency signals will also induce the extra error,  which can be extracted refer to (\ref{Eq:18}):

\begin{equation}
\rm{Error}=\rm{FFV}_{PCNN}-\rm{FFV}_{TCNN}-\rm{Weighting\,Error}
\label{Eq:18}
\end{equation}
where \(\rm{Weighting\,Error}\) is fixed at any baud rate in our simulator. We run the photonic CNN at the baud rate of 5, 10, 15, 20, and 25 Gbaud for 10 samples. The distribution statistics of \(\rm{Error}\) with 360 elements at each baud rate is shown are Fig. \ref{fig:7} (a) to (e). 
To analyze the impact of levels of error  on the test accuracy at different baud rates, the probability density function (PDF) of  the error at each baud rate are calculated. 
The PDF shows a normal distribution, and the Gaussian fit curve of PDF at each baud rate is shown in Fig. \ref{fig:7}(f). 
The mean value of Gaussian fit function will decrease whereas variance increases at higher baud rate for input vector, meaning that the error will increase with the baud rate. 
10 random error sequences \({\rm{Error}}'_i\) are generated according to the PDF at each baud rate and added with \(({\rm{FFV}}_{TCNN}+{\rm{Weighting\,Error}})\), which are combined as new flatten feature vector with errors sent to the classifier for testing.
The performance of photonic CNN at different baud rate is shown in Fig. \ref{fig:8}. Note that the distance between the optimal point and the convergence point is shown in Fig. \ref{fig:6}(d). 
The difference of average accuracy at each baud rate and standard deviation of test accuracy should be considered instead. In Fig. \ref{fig:8}, the performance  degrades with the increasing of baud rate, showing that the high speed photonic CNN will pay its the cost of computation performance.  However, high operation baud rate will mean less computing time, which can be roughly calculated as
\begin{equation}
t_{2Dconv}=[M\times (M+2)+2]/BR+t_c
\label{Eq:19}
\end{equation}
\begin{figure}[h]
\begin{center} 
\includegraphics [width=0.75\textwidth]{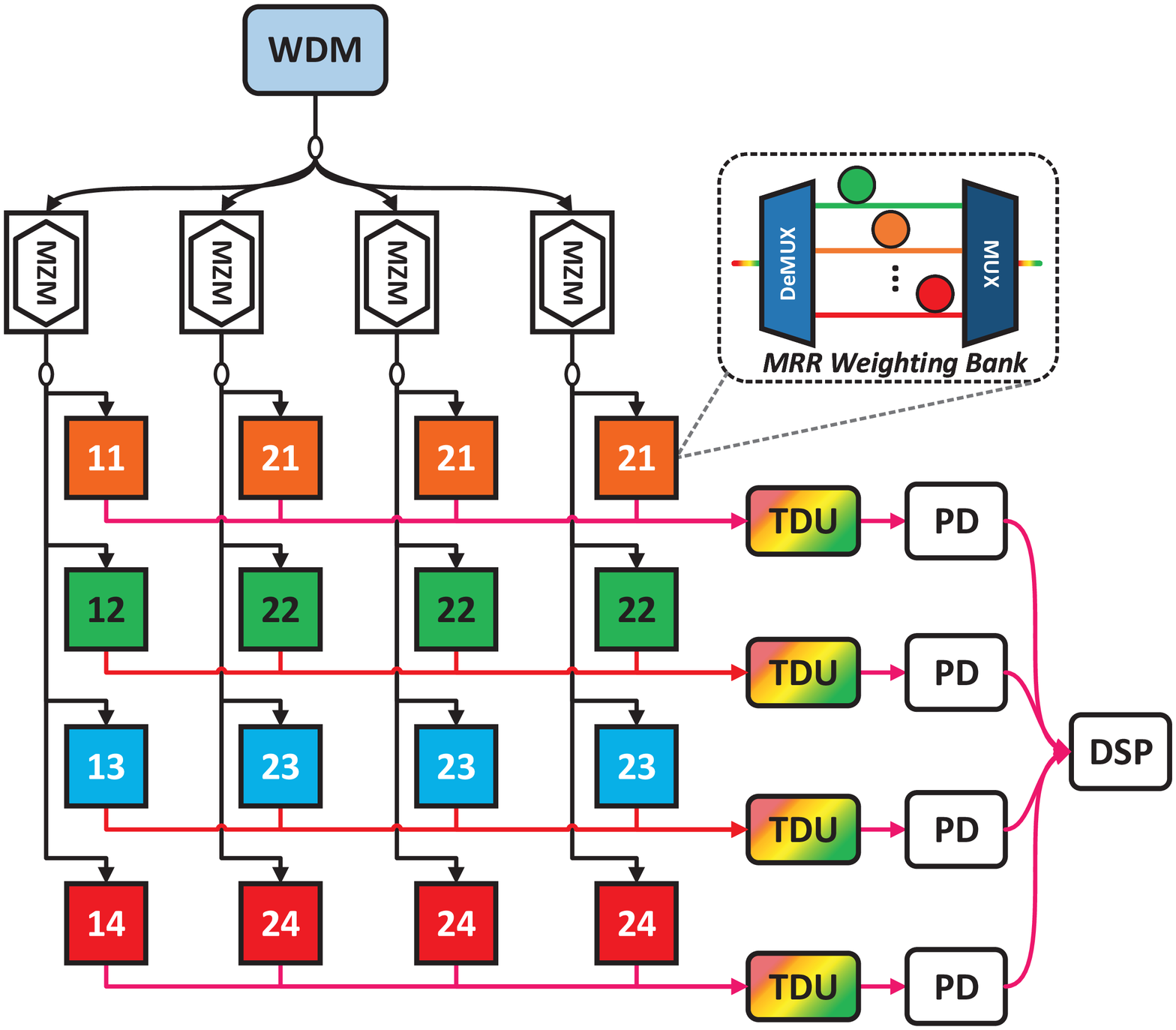} 
\caption{PCNN based on \(\mathcal{C}\times \mathcal{K}\) MWB (MRR weighting bank) mesh, the MWB in each column are for the same input channel in different kernels, and the MWB in each row combine as one kernel with \( \mathcal{C}\) channels. }  
\label{fig:9}
\end{center}
\end{figure}

\begin{table*}[t]

\setlength{\extrarowheight}{1pt}
\begin{center}
\setlength{\abovecaptionskip}{-10pt}
\caption{EXECUTION SPEED AT DIFFERENT BAUD RATE FOR PCNN WITH 1 OCU}
\label{Tab:1}
\resizebox{\textwidth}{20mm}{
\begin{tabular}{cccccccc}
\toprule
\begin{tabular}[c]{@{}c@{}}Baud \\ Rate\end{tabular} & \begin{tabular}[c]{@{}c@{}}Time of Conv.1\\ (M=28)\\ Period =2\end{tabular} & \begin{tabular}[c]{@{}c@{}}Time of Conv.2\\ (M=13)\\ Period =8\end{tabular} & \begin{tabular}[c]{@{}c@{}}Time of Conv.3\\ (M=5)\\ Period=16\end{tabular} & \begin{tabular}[c]{@{}c@{}}Total\\ time\end{tabular} & Ops                    & \begin{tabular}[c]{@{}c@{}}Execution Speed\\ (Average)\end{tabular} & \begin{tabular}[c]{@{}c@{}}Execution Speed\\ (2Dconv)\end{tabular} \\ 

\midrule
5G & 340 ns & 320 ns   & 128 ns & 788 ns    &   & 56 GOPS   & 71 GOPS\\
10G& 170 ns  & 160 ns  & 64 ns  & 394 ns      &      & 112 GOPS     & 143 GOPS   \\
15G & 114 ns & 112 ns  & 40 ns  & 266 ns   & 44352  & 166 GOPS      & 213 GOPS   \\
20G & 86 ns  & 80 ns   & 32 ns  & 198 ns    & & 224 GOPS  & 282 GOPS    \\
25G & 68 ns  & 64 ns   & 24 ns   & 156 ns  &  & 284 GOPS  & 357 GOPS   \\ 
\bottomrule
\end{tabular}
}
\end{center}
\end{table*}
\begin{table*}[h]
\begin{center}
\setlength{\abovecaptionskip}{-10pt}
\caption{EXECUTION SPEED AT DIFFERENT BAUD RATE FOR \(4\times 4\) PCNN MESH}
\label{Tab:2}
\resizebox{\textwidth}{18mm}{
\setlength{\extrarowheight}{1pt}
\begin{tabular}{cccccccc}
\toprule
\begin{tabular}[c]{@{}c@{}}Baud \\ Rate\end{tabular} & \begin{tabular}[c]{@{}c@{}}Time of Conv.1\\ (M=28)\end{tabular} & \begin{tabular}[c]{@{}c@{}}Time of Conv.2\\ (M=13)\end{tabular} & \begin{tabular}[c]{@{}c@{}}Time of Conv.3\\ (M=5)\end{tabular} & \begin{tabular}[c]{@{}c@{}}Total\\ time\end{tabular} & Ops & \begin{tabular}[c]{@{}c@{}}Execution Speed\\ (54$\%$ Utilization)\end{tabular} & \begin{tabular}[c]{@{}c@{}}Execution Speed\\ (100$\%$ Utilization)\end{tabular} \\ 
\midrule
5G & 170 ns & 40 ns & 8 ns & 218 ns &  & 203 GOPS & 324 GOPS \\
10G & 85 ns & 20 ns & 4 ns & 109 ns &  & 406 GOPS & 648 GOPS \\
15G & 57 ns & 14 ns & 2.5 ns & 73.5 ns & 44352 & 603 GOPS & 1.03 TOPS \\
20G & 43 ns & 10 ns & 2 ns & 55 ns &  & 806 GOPS & 1.29 TOPS \\
25G & 34 ns & 8 ns & 1.5 ns & 43.5 ns &  & 1.02 TOPS & 1.73 TOPS \\ 
\bottomrule
\end{tabular}}
\end{center}

\end{table*}
 Where \(t_c\) is the time delay in \(\rm{OM^2U}\), which is usually less than 100 ps in our system. Thus, the execution speed at different are as shown in Table \ref{Tab:1}.
 Note that the operation in TCNN is a 4-dimension operation (or tensor operation) for width, height, channel and kernel. However, for each OCU only 2-dimension operation for width, height is realized during one period. In the layer of a photonic CNN with input of \(\mathcal{C}\) channels and \(\mathcal{K}\) kernels, one OCU can be used repeatedly to complete 4-dimension operation in \(\mathcal{C}\times \mathcal{K}\) periods.  
 To improve the execution speed, the parallelization of the photonic CNN is necessary in the future. 
 In this paper, a candidate mesh with MRR weighting bank shown in Fig. \ref{fig:9} is proposed to complete tensor operation during one period. 
 Each row of the mesh is combined as one kernel with all channels. And the same channel of input figure is copied and sent to the mesh in the same column. 
 For the first layer of photonic CNN, the input image “8” is flattened into \(1\times 784\) vector and duplicated into two copies by a splitter for \(MWB_{1,1}\) and \(MWB_{2,1}\).
 Two \(1\times 842\) vectors are sent to the DSP through the TDU and PD in the \(1^{st}\) and \(2^{nd}\) row of mesh. 
 Note that the length of optical path through mesh and dispersion media should be equal.
 The execution speed of the \(4\times 4\) mesh at different baud rate is shown in Table. \ref{Tab:2}.
 Note that the mesh is not 100$\%$ utilized in each period when loaded a simplified AlexNet shown in Fig. \ref{fig:4}(a). 
 The average utilization of PCNN can be calculated as \(2/16+8/16+16/16=54\)$\%$, thus the average execution time for one sample is much lower due to nature of parallelization.
 Refer to (\ref{Eq:16}) and Table \ref{Tab:1}, and \ref{Tab:2}, the photonic CNN running at higher baud rate has faster execution speed and lower delay scale.
 However, the selection of baud rate depends on  
 the requirement of CNN performance and time delay resolution.
 As shown in Fig. \ref{fig:8}, the performance degenerate significantly at \(Baud\,Rate=25\) G. 
Moreover, if we choose the delay structure in Fig. \ref{fig:3}, and we set the length of DCF of \(L=2 km\) and comb with density of \(0.2\) nm, \(\mathcal{R}= 60\) ps according to (\ref{Eq:16}), which allows \(Baud\,Rate \leq 16.7\) G.

\begin{figure}[h]
\begin{center} 
\includegraphics [width= 0.9\textwidth]{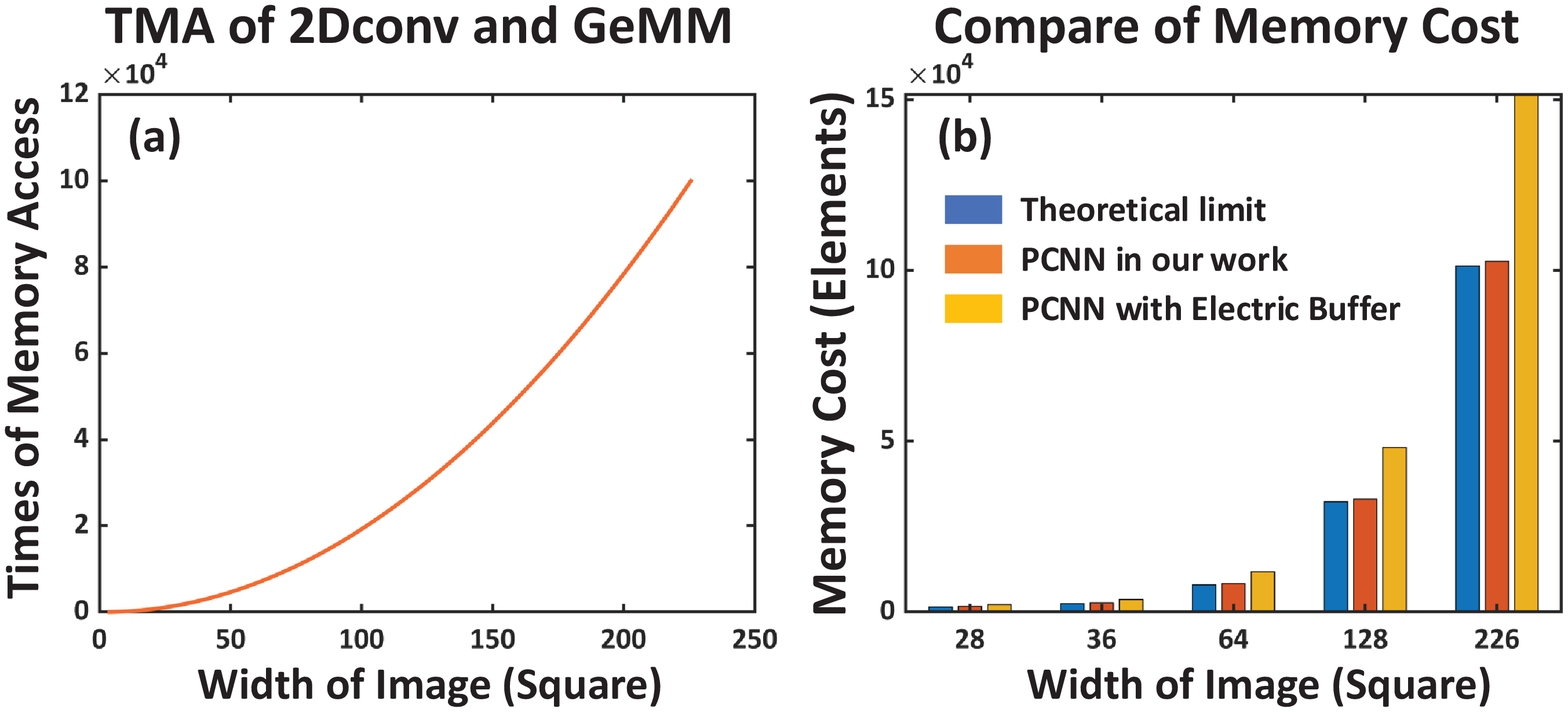} 
\caption{(a) Times of Memory Access (TMA) of 2Dconv, GeMM algorithms and the PCNN based on electronic buffer, where as the memory access of the PCNN based on TDU is fixed and equal to 2, (b) Compare of Memory cost between PCNN in ourwork, PCNN based on electric buffer and 2Dconv algorithm, which is also the theoretical lower limit of CNN.}  
\label{fig:10}
\end{center} 
\end{figure}

\subsection{Memory Cost}
The photonic CNN using electronic buffer based on 2Dconv and GeMM algorithm need to access to memory reapeatly to extract the corresponding image slice.
The number of times for memory access is \(2\times (M-N+1)^2\).
As shown in Fig. \ref{fig:10}(a),  memory access times for 2Dconv and GeMM algorithm will increase significantly with the width of input image, since that multiplication, addition and zero padding operations will require a large amount of data in memory shown in Fig. \ref{fig:10}(b).
However, photonic CNN only needs to take out the flatten image vector and store the convolution results, i.e. only 2 times for memory access are needed. Further more, intermediate data stored in the optical delay unit which will have less memory cost compared to electrical counterpart as in Fig. \ref{fig:10} and very close to the theoretical lower limit.

\section{Conclusion}
In this paper,  we propose a novel integrated photonic CNN
based on double correlation operations through interleaved timewavelength modulation.
200 images are tested in MNIST datasets with accuracy of 85.5$\%$ in our PCNN versus 86.5$\%$ in 64-bit computer.
The error caused by distortion induced by filters and ADC will increases with the baud rate of the input images, leading to the degradation of classification performance.
A tensor processing unit based on \(4\times 4\) mesh  with 1.2 TOPS (operation per second when 100$\%$ utilization) computing capability at 20G baud rate is proposed and analyzed to form a paralleled photonic CNN.

\bibliographystyle{unsrt}  

\bibliography{references}  






\end{document}